\documentstyle[12pt]{article}

\begin{document}
\topmargin-10mm
\oddsidemargin-5mm

\setcounter{page}{1}
%\begin{titlepage}
\rightline{Preprint YRPHI-1995(6)-95}
\vspace{1cm}
\begin{center}

{\bf{COAXIAL RING CYCLOTRON AS A PERSPECTIVE\\
NUCLEAR POWER ENGINEERING MACHINE}}\\
\vspace{5mm}
{\large A.R.Tumanian, Kh.A.Simonian, R.L.Mkrtchian,
A.Ts.Amatuni,\\
R.O.Avakian}\\
\vspace{5mm}
{\em Yerevan Physics Institute}\\
{\large A.G.Khudaverdian}\\
{\em Yerevan State University}\\
\end{center}

\vspace{5mm}
\centerline{{\bf{Abstract}}}
The  circuit  arrangement  of  the  proposed  coaxial   ring
cyclotron (CRC) is described, and its main  advantages,  such  as
simple injection technique, several  injected  beams  summation
option, high efficiency,  are  considered.  The  proposed  proton
accelerator is a perspective machine for the solution of the main
problems of the present day nuclear power engineering as well  as
for the next-generation  nuclear  power  plants,  representing  a
combination of subcritical reactors  and  particle  accelerators.
The possibility of installation of CRCs  into  ring  accelerators
with an average diameter from  60  to $100m$,  e.g.,  the  Yerevan
electron synchrotron, is considered.

\vspace{5mm}
\centerline{\bf Introduction}
\indent
In recent years a complex  of  major  applied  problems  has
clearly  crystallized,  for  successful  solution  of  which  are
required powerful proton accelerators with proton beam  intensity
from 5 to $100mA$, energy from 500 to 800 MeV, and high  efficiency.
To  such  problems  relate,   e.g.,   production   of   different
radioactive isotopes, turning nuclear  fuel  out  from $U^{238}$  and
Th$^{232}$, nuclear power plant radioactive waste treatment, designing
of safe  (subcritical)  and  next-generation  waste-free  nuclear
power plants, which can utilize unenriched uranium or thorium. As
well as in this connection, due to the successful  conduction  by
C. Rubbia's group,  CERN  [1]  of  the  experiment  approving  in
principle a new and  perspective  possibility  of  nuclear  power
production using proton beams  and  a  subcritical  reactor,  the
development  of  appropriate  accelerators  becomes   an   urgent
necessity.

\vspace{5mm}
\centerline{\bf Comparison of Known-Type Accelerators}
\indent
Leaving the consideration of  new  acceleration  techniques,
let us dwell on traditional machines, such  as  $cw$
linacs  and  isochronous  cyclotrons,  suitable  for  the   goals
mentioned. Linac projects are developed mainly in the USA,  while
isochronous cyclotrons in West  Europe  and  Russia,  the  reason
being probably connected with the achievements of these countries
in the relevant fields of accelerator engineering.

A comparative analysis of these two  types  of  accelerators
has been done in [4,5]. The main drawback of linacs is that for a
given beam the accelerating system is used only once,  which  is
not the case with a cyclotron, where it is used many times.  This
inevitably increases sharply the power of  losses,  which,  e.g.,
for such mesonic factories reaches $40-50MW$ in pulse [2]. In  case
of continuous operation of such linacs, the power of losses  will
essentially affect the overall efficiency which is one  of  the
decisive
factors in designing commercial  models.  There  are  also  other
considerations discussed in details in [3,4,5].

To the drawbacks of the cyclotron-type  accelerators  relate
the complexity of injection and summation of several beams, and
the earlier problem of $a 100\%$  extraction  of  powerful  megawatt
beams, which caused in some projects the preference to  be  given
to linear accelerators [5]. Later, however, the  beam  extraction
problem  was  successfully  resolved  theoretically  [6]  with  a
following  experimental  verification   [7]   realized   in   the
isochronous  cyclotron  SIN-PSI  VILLIGEN,  Switzerland,   having
record parameters  to  the  present  day: $E=590$MeV, $I=1.0-1.5mA
[3,13]$.  But  the  problem  of  several  beams  summation   and
injection into a cyclotron still remains open. By  the  way,  the
problem of summation of, to our regret, only two beams has been
successfully resolved for linacs, by creating the "FUNNEL" system
suggested by B.Montague and  K.Bongardt  [8,9],  which  has  been
tested at LANL [10]. In the present paper is considered a  method
of solution of that problem for cyclotrons, proposed  by  one  of
the coauthors (Tumanian A.R.).

\vspace{5mm}
\centerline{\bf Coaxial Ring Cyclotron}
\indent
It is known, that the proton beam  injection  path  usually
passes from above or from below of the  magnetic  system  of  any
type of cyclotrons. Then with the help  of  a  group  of  bending
magnets, the protons are transferred onto the  median  plane  and
then with the help of a  system  of  magnetic  and  electrostatic
deflectors are injected into the ring-shaped chamber. In  such  a
bulky arrangement, it is problematic to organize  a  simultaneous
injection and summation of two ormore beams.

It is suggested to increase the diameter of the  isochronous
ring cyclotron so, that it  becomes  possible  to  place  in  its
central part several injectors  (cyclotron  or  other  type,  see
Fig.1) in a coaxial  arrangement.  At  such  an  arrangement,  it
obviously becomes easier to inject several beams into the  same
ring, into different by radius orbits. So, let us  name  the  new
method proposed as "orbital" method of filling the ring  up  with
beams.

Orbital filling of the cyclotron ring  is  provided  by  the
fact, that the energy of each injector differs from that  of  the
next one by $\Delta E^{i}_{\hbox{inj}}$, the  conditions  of
separation of  different injected bunch orbits being met:
$$
\sum^{m}_{i=1} \Delta E^{i}_{\hbox{inj}} < \Delta E_{\hbox{cycl}} ;
\qquad m[A+r(\Delta E/E)] < \Delta R_{\hbox{cycle}},
$$
\noindent where $r(\Delta E/E)$ is  "orbit  width",  A  is  amplitude
of  betatron
oscillations, $\Delta R_{\hbox{cycl}}$  is radial gain per turn,
$\Delta E_{\hbox{cycl}}$  is
energy gain per turn, $\Delta E/E$  is  energy
spread, $m$ is number of injectors. All "$m$" beams can be  extracted
from one or different outlets on the ring, the number of  outlets
not exceeding the number "$m$".

The average radius of the final orbit is chosen with account of
the following considerations. On  the  one  hand,  the  machine's
diameter as well as the magnetic poles  width  do  increase  with
orbit's radius, this increasing capital outlays. But on the other
hand,  larger  radius  leads  to  desirable  decreasing  of   "$B$"
induction in  cyclotron  magnets,  this  making  it  possible  to
exclude  the  main  power  supply  system  of  magnets.  Such   a
possibility arises at less than 3.0 kGauss induction in magnets.

In this case, it is possible to use cast  permanent  magnets
made of hard magnetic materials  (Alnico,  Magnico,  etc.).  Such
magnets have practically unrestricted forms  and  sizes  and  are
widely used in powerful klystrons, magnetrons, etc. [12].

Important is the fact, that  exclusion  of  the  main  power
supply system of  magnets  leads  to  an  obvious  and  essential
improvement of cyclotron overall efficiency, as the power of  such  power
supply systems makes hundreds of kilowatts, e.g., $650kW$  for  SIN
(at $E=590$MeV, $I=0.1mA)$ at a total  power  consumption  of $1500kW
[13]$.  So,  at  exclusion  of  the  power  supply   system,   the
cyclotron's efficiency will improve more than 1.7 times
, this essentially shortening the pay-back period.

\vspace{5mm}
\centerline{\bf The YerPhI Version of CRC}
\indent

The  basic  parameters  of  the  Yerevan  Physics  Institute
version of CRC are: $R_{m}=34.5m$; tunnel width $- 6.0m$;  thickness  of
tunnel  walls $- 3.0m$;  inner  diameter  -  57m.  Six   SIN-type
cyclotrons  serve  as  injectors,   each   with $15m$   diameter,
$E_{\hbox{inj}}=500$MeV, I=1.5mA. Assuming,  that  CRC  has  48
acceleration
gaps, 32 of which are occupied by SIN-type resonators, the energy
gain  per  turn  will  make $\Delta E_{\hbox{cycl}}=16$MeV,
the radial gain per turn at the injection level making $\Delta
R^{\hbox{inj}}_{\hbox{cycl}}\cong 26$cm. The
average radius $R_{500}=32.5m$ for particles with
$E_{\hbox{inj}}=500$MeV,  and
the  working  area  (magnetic  poles  width)  should  not  exceed
$\Delta R_{\hbox{mag}}=3.7$m. Then it is easy to notice,
that in such a ring with a
final radius of $R_{\hbox{fin}}=36.1m$ the protons can be
accelerated  up  to
$E_{\hbox{fin}}=800$MeV. The average value of induction  in  magnets
will  vary with energy between $1.117\div 1.352$kGauss.

Note, that the form of magnets  and  the  field  topography,
spirality, variation depth, etc. should be defined  with  account
of axial betatron oscillations stability conditions.

Suppose  the  injected  beam  radial  size,  equal  to  the
amplitude of betatron oscillations  in  CRC, $A=1cm$,  and  energy
spread $\Delta E/E_{\hbox{inj}}=10^{-3}$, then from six different SINs, there
can  be
injected into the cyclotron magnetic ring beams with an  energy
increment of $\Delta E^{i}_{\hbox{inj}}=2.5$MeV, providing  a  complete
separation  of
beams.  The  value  of $\Delta E^{i}_{\hbox{inj}}$  will   remain   unchanged
at   a
correlational change  of  beam  injection  parameters,  e.g.,  at
$A=0.5cm$ and $\Delta E/E_{\hbox{inj}}=2\times 10^{-3}$.
\par
Simple calculations show,  that  with  the  particle  energy
increasing from 500 to 800MeV, the radial gain per turn
decreases from $\Delta R^{\hbox{inj}}_{\hbox{cycl}}=26.2cm$
down  to $\Delta R^{\hbox{inj}}_{\hbox{cycl}}=13.8$cm.
Nevertheless, all six beams will be spaced,  as  far  as  in  the
process of acceleration, both the betatron oscillation  amplitude
and energy spread decrease, this narrowing the  regions  occupied
by beams.

On the basis  of  the  PSI  cyclotron  update  results,  the
global efficiency of the three-stage cyclotron complex  (regular  900MeV
machine, $I=10mA)$ is estimated to be $40\% [14]$. Assuming  the  same
efficiency as minimum for the YerPhI version of CRC combined with
a subcritical thorium reactor  [14],  it  would  be  possible  to
produce a useful electrical power of more than 70MW.

\vspace{5mm}
\centerline{\bf Conclusion}
\indent

The  scientific  and  technical  solutions  and  preliminary
estimations, the results of which are presented in  this  report,
would be sufficient to start numerical calculations of  different
versions, simulation of the basic systems and units, and  develop
a tentative CRC version. The purpose of the present report is  to
show there are no restrictions  for  creation  of  a  CRC,  which
economically prevails over its counterparts.

\vspace{5mm}
\centerline{\bf Acknowledgements}
\indent

The  authors  thank  M.L.Petrossian,   S.G.Arutunian,   V.M.
Tsakanov, I.P.Karabekov for detailed and helpful discussions  and
valuable remarks.

\centerline{\bf References}
\indent

\noindent 1. Europhys News, 26 (1995); CERN Courier No.3 (1995).

\noindent 2. D.C.Hagerman, IEEE Transaction on Nuclear Science,
No.5-13(4), p.277, 1966.

\noindent 3. Uwe Trinks, EPAC-94, v.1, p.295.

\noindent 4. V.P.Djelepov et al., Preprint JINR $P-9-9066,$ Dubna,
1975  (in Russian).

\noindent 5. G.A.Barthalamew et al., AECL-2750, 1967.

\noindent 6. V.P.Dmitrievski et al., JINR Rep. $P-9-6733,$  Dubna,
1972  (in Russian).

\noindent 7. A.T.Vasilenko et al., Proc. 4th All-Union Conf.,  v.1,
p.205, Moscow, 1975 (in Russian).

\noindent 8. B.Montague, CERN Rep., K.John
son, Priv. comm.

\noindent 9. K.Bongardt, D.Sanitz, HIIF, GST 87-8 (1972), 224.

\noindent 10.K.F.Johnson et al., Linac$-90, LA-12004-C (1990)$.

\noindent 11. L.A.Sarkisian, Proc. $2nd$ All-Union Conf., v.1,  p.33,
Moscow, 1970 (in Russian).

\noindent 12."Elektronika" encyclopedic dictionary, Moscow, 1991.

\noindent 13.H.A.Willax et  al.,  Proc $2nd$  All-Union  Conf.  on
Particle Accelerators, Moscow, 1970.

\noindent 14.C.Rubbia, EPAC-94, v.1, p.270.

\end{document}